\documentclass[sigconf]{acmart}

\AtBeginDocument{}

\copyrightyear{2025}
\acmYear{2025}
\setcopyright{rightsretained}
\acmConference[CUI '25]{Proceedings of the 7th ACM Conference on
Conversational User Interfaces}{July 8--10, 2025}{Waterloo, ON, Canada}
\acmBooktitle{Proceedings of the 7th ACM Conference on Conversational User
Interfaces (CUI '25), July 8--10, 2025, Waterloo, ON, Canada}
\acmDOI{10.1145/3719160.3737632}
\acmISBN{979-8-4007-1527-3/2025/07}

\begin{document}

\title{Fitting the Message to the Moment: Designing Calendar-Aware Stress Messaging with Large Language Models}

\author{Pranav Rao}
\authornote{Rao and Taj contributed equally to this work. Their names are listed in alphabetical order by last name.}
\affiliation{%
  \institution{Computer Science, University of Toronto}
  \city{Toronto}
  \state{Ontario}
  \country{Canada}
}

\author{Maryam Taj}
\authornotemark[1]
\affiliation{%
  \institution{Computer Science, University of Toronto}
  \city{Toronto}
  \state{Ontario}
  \country{Canada}
}
\author{Alex Mariakakis}
\affiliation{%
  \institution{Computer Science, University of Toronto}
  \city{Toronto}
  \state{Ontario}
  \country{Canada}
}

\author{Joseph Jay Williams}
\affiliation{%
  \institution{Computer Science, University of Toronto}
  \city{Toronto}
  \state{Ontario}
  \country{Canada}
}

\author{Ananya Bhattacharjee}
\affiliation{%
  \institution{Computer Science, University of Toronto}
  \city{Toronto}
  \state{Ontario}
  \country{Canada}
}

\newcommand{\todo}[1]{\textcolor{red}{\{TODO: #1\}}}
\renewcommand{\shortauthors}{Rao et al.}
\newcommand{\italquote}[1]{\begin{quote}``\textit{#1}''\end{quote}}
\begin{abstract}
Existing stress-management tools fail to account for the timing and contextual specificity of students’ daily lives, often providing static or misaligned support. Digital calendars contain rich, personal indicators of upcoming responsibilities, yet this data is rarely leveraged for adaptive wellbeing interventions. In this short paper, we explore how large language models (LLMs) might use digital calendar data to deliver timely and personalized stress support. We conducted a one-week study with eight university students using a functional technology probe that generated daily stress-management messages based on participants’ calendar events. Through semi-structured interviews and thematic analysis, we found that participants valued interventions that prioritized stressful events and adopted a concise, but colloquial tone. These findings reveal key design implications for LLM-based stress-management tools, including the need for structured questioning and tone calibration to foster relevance and trust.

\end{abstract}

\ccsdesc[500]{Human-centered computing~Empirical studies in HCI}

\keywords{Stress Management, Digital Calendars, Large Language Models, Text Messaging}

\maketitle

\section{Introduction}
Stress is commonly defined as a psychological or emotional strain triggered by demanding circumstances \cite{lazarus1984stress}. University students are particularly vulnerable, facing a combination of academic pressures, shifting responsibilities, financial concerns, and personal transitions \cite{duffy2020predictors, barbayannis2022academic, moore2021qualitative}. Prolonged stress in this population has been linked to serious psychological and physiological outcomes, including anxiety, depression, and cardiovascular disease \cite{beiter2015prevalence, doi:10.1161/CIRCIMAGING.120.010931}. In response, numerous digital tools have been developed to support student wellbeing, offering strategies for relaxation, emotional regulation, and time management. While these interventions have shown promise \cite{nahum2018just, kornfield2022involving}, a recurring limitation is their inability to adapt to the situational complexity of students' lives, often leading users to perceive them as generic or misaligned with their actual needs \cite{bhattacharjee2023design}.

Digital calendars offer a valuable but underutilized source of contextual information about students' ongoing tasks and responsibilities \cite{howe2022design}. Many students use calendars extensively \cite{mei2016learning}, and the event details they input—titles, descriptions, and notes—capture meaningful, real-world pressures. However, this information is typically unstructured and difficult for traditional support tools to process. Advances in large language models (LLMs) offer a promising solution, as LLMs are capable of interpreting personalized and natural language content. Building on emerging research that uses LLMs to support wellbeing in educational settings \cite{bhattacharjee2024understanding, kumar2024large, bhattacharjee2024explains, rao2024integrating}, we see an opportunity to connect LLMs and calendars to deliver adaptive, situation-specific stress support. Yet, this design space remains largely unexplored.

Motivated by these opportunities and challenges, we investigate how students envision the role of LLMs in supporting stress management through their digital calendar events. Rather than designing full systems, we first aim to understand how students imagine LLMs could deliver more personalized and adaptive coping strategies. Specifically, we ask:

\begin{itemize} 
\item \textbf{RQ:} How do university students envision the role of LLMs in leveraging calendar data to tailor strategies for managing stress? 
\end{itemize}

To answer this research question, we conducted a one-week study with eight university students experiencing at least moderate levels of stress. Participants used a functional technology probe \cite{hutchinson2003technology} that delivered daily text messages for stress support. Messages were adapted from existing content using an LLM to reflect participants’ digital calendar events. We found that participants expressed a desire to prioritize messages about stressful events and maintain a concise yet colloquial tone.  This study contributes to human-computer interaction by examining early user expectations and responses to the integration of digital calendar data with LLM-generated, personalized stress management interventions.

\section{Related Work}




Prior work has shown that personalization of stress-management interventions based on contextual factors can greatly increase their effectiveness \cite{doherty2012engagement, paredes2014poptherapy, howe2022design, bhattacharjee2023investigating, nepal2024contextual}. Timing is one such factor: just-in-time adaptive interventions, which aim to deliver support when users need it most, have been found to outperform randomly timed alternatives \cite{nahum2018just, smyth2016providing}. Several studies have explored using digital calendars to inform time-sensitive wellbeing interventions \cite{howe2022design, kocielnik2013smart, tateyama2022mood}. For instance, Howe et al. \cite{howe2022design} found that interventions aligned with Outlook calendar events increased engagement and reduced stress, while Kocielnik et al. \cite{kocielnik2013smart} used calendar and sensor data to help users visualize and reflect on long-term stress patterns. Although this work highlights the value of digital calendars for identifying stress patterns and informing timing, it does not leverage the rich semantic content of the calendar events themselves. Our design introduces LLMs to interpret this content and support deeper personalization.

By tailoring content to a user’s situation and emotional state, LLMs can make interventions feel more relevant and personally meaningful \cite{kumar2024large, bhattacharjee2024explains, bhattacharjee2024understanding}. Recent studies have begun to explore the affordances of connecting LLMs with digital calendars for planning and scheduling \cite{hoefer2025telltime, gundawar2024robust, saxena2025lost}. For example, Hoefer et al. used LLMs to transform spoken narratives into structured calendar entries, allowing users to reconstruct their day with less manual effort \cite{hoefer2025telltime}. However, little is known about how such integrations might support stress management. We take an initial step towards addressing this gap by examining how students envision the role of LLMs in providing context-aware stress support using calendar data.



\section{Design of the Probe}
\begin{figure*}[ht]
    \centering
    \includegraphics[width=0.9\linewidth]{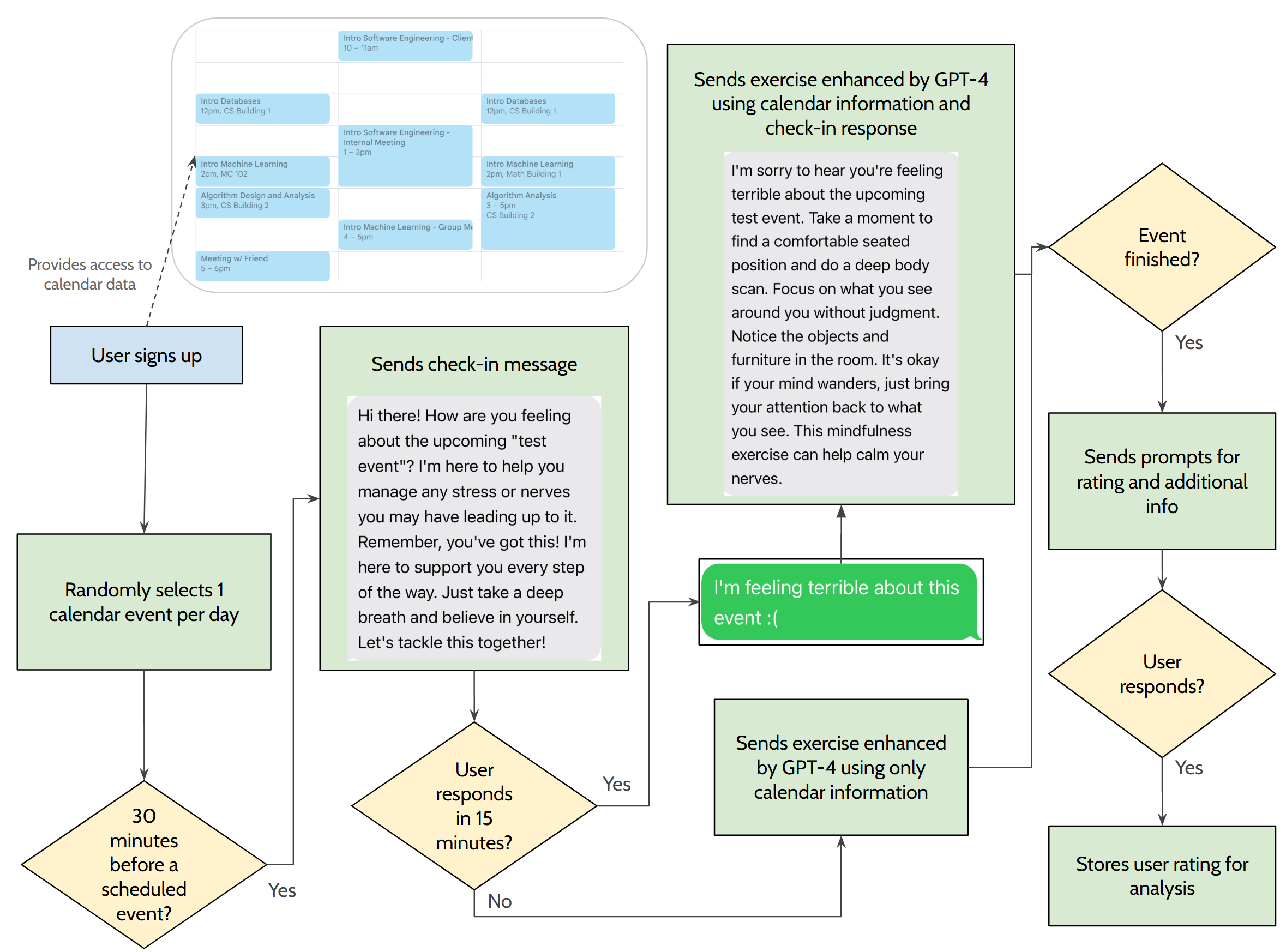}
    \caption{Daily flow of the calendar-based LLM probe}
    \label{fig:fig-1}
\end{figure*}

We developed a lightweight interactive probe that delivered daily text messages tailored to the timing and content of participants’ Google Calendar events. The design of this probe was inspired by Rao et al. \cite{rao2024integrating}. Importantly, the probe was not designed to function as a fully adaptive intervention system, but rather to elicit participants’ expectations and reactions to this type of support. Each day, the system randomly selected an event from the participant’s calendar. Shortly before the event's scheduled time, it prompted them to report how they were feeling and used GPT-4 to select and adapt a message from a curated set of expert-authored stress-reduction exercises \cite{bhattacharjee2023design, meyerhoff2022system}. GPT-4 received contextual information from the calendar and, when applicable, the participant’s check-in response. After each event, participants were asked to rate and briefly comment on the helpfulness of the message. A visual overview of the system’s message flow is shown in Figure \ref{fig:fig-1}.

\section{Method}
\subsection{Participants}

We recruited eight participants (6 women, 2 men) from a large North American post-secondary institution through snowball sampling. Participants were invited to the study through flyers distributed by members of the research team. Participants were required to be between the ages of 18 and 25 and reside in North America. The mean age of our participants is 20.4±0.2 years old, and they identified with several racial groups (5 Asian, 1 White, 1 African, 1 Mixed-Race). Additionally, they had to self-identify as at least moderately stressed, scoring 14 or higher on the Perceived Stress Scale (PSS) \cite{cohen1983pss}. After agreeing to participate, individuals completed the PSS via a provided link and notified the research team. Eligible participants were enrolled based on their scores. All participants self-identified as frequent users of both digital calendars and LLMs. Throughout the one-week probe trial, a research team member continuously monitored the system's messages and participant responses to ensure correct system functionality. We refer to the participants as P1–P8.

\subsection{Procedure}
Our research procedure was approved by the Research Ethics Board at the University of Toronto. After the study period, we collected feedback through one-on-one, semi-structured interviews conducted via Google Meet. Each session, facilitated by two members of the research team, lasted between 30 and 50 minutes. Participants first received a brief overview of the study’s objectives and a walkthrough of the text messages they had received from the application. Participants were then invited to share their thoughts on the utility of the tool and potential features they would find beneficial. Interview questions included, but were not limited to, "Did you complete the suggested stress management exercise? Why or why not?" and "Imagine you could customize the way LLM interacts with you. What aspects would you want to change?" Finally, all participants received \$10 USD compensation for their involvement in the study. 

\subsection{Data Analysis}

The interviews yielded approximately 3.8 hours of audio recordings corresponding to 8 transcripts. We applied a thematic analysis approach to extract relevant themes from the collected data. To structure the analysis, we organized each transcript by interview question and conducted coding at the level of individual responses. Two team members independently created initial codebooks by openly coding one shared transcript using this question-based structure. They then employed a consensus coding approach on this transcript, as described by Braun and Clarke \cite{braun2006thematic}, engaging in collaborative sessions to compare interpretations, reconcile differences, and refine the codebook to reflect the research questions. Once finalized, they applied the codebook to the remaining transcripts. Using axial coding, we grouped related codes into broader themes, each supported by multiple excerpts from the data. These excerpts provided concrete evidence for each theme, strengthening the validity of our analysis and shaping the organization of our discussion. Throughout the coding process, we revisited transcripts to ensure consistency, coherence, and alignment with participant perspectives.

\subsection{Ethical Considerations}
Given the focus on stress management, we were mindful of the ethical considerations involved in deploying AI-generated content in emotionally sensitive contexts. Participants were informed that this was an LLM-enhanced technology probe, which could carry a low risk of producing content that felt misaligned, impersonal, or unintentionally triggering. We reviewed all incoming and outgoing messages daily and were prepared to intervene or follow-up if any adverse reactions were observed, similar to prior studies \cite{bhattacharjee2022kind, bhattacharjee2023design}. However, no such interventions were necessary during the study. To further safeguard participant wellbeing, we made it clear at the outset of the study and during interview sessions that participants could skip any questions or withdraw at any point without consequence.

\section{Results}
Participants interacted with the system during 44\% of recorded events. However, it is important to note that the probe's purpose was not to maximize engagement but to expose participants to a system that integrates calendar data to deliver contextual stress support over a week. This exposure allowed participants to reflect on their experiences and provide informed feedback.
In this section, we present key themes from participant interviews, including their perceptions about the system's selection and interpretation of events and message style.

\subsection{Challenges in Event Selection and Interpretation}
Participants found the system most engaging when it selected events they perceived as stressful. However, the system frequently sent messages for routine or non-stressful events. As P1 explained, these messages felt unhelpful because, \textit{"I'm receiving unsolicited advice on what to do with this event when, like, I never sort of asked for it in the first place."} Furthermore, receiving messages for non-stressful events occasionally induced unintended anxiety or stress. P2 illustrated this, saying, \textit{"Let’s say I was only feeling excited about this [event], and then I see this text, and then it asks whether I am feeling excited, anxious or a mix of both. And then, I’m like, maybe, I am anxious about it."}

A few participants also reported that the system occasionally misinterpreted their events. For example, P3 recalled that the system provided an exercise for performing at a show, when they were attending instead. They attributed this to vague or minimal calendar entries (i.e., using acronyms or shorthand phrases). As P4 said, \textit{"even maybe a person would have trouble understanding what I mean by that."}

Despite these instances, the messages were appreciated by some, even for routine events, as they introduced more intentionality into their day. P8 shared, \textit{“It did prompt me to pause for a moment. So, it was very helpful in that sense.”} These participants also noted that the messages fostered greater awareness about their commitments, promoting productive reflection on how they spent their time and energy.

\subsection{Reduced Engagement with Longer Messages and Perceived Artificiality}
Participants responded most positively to messages that ended with a clear call to action, as this provided a tangible next step and a sense of purpose. Conversely, engagement noticeably decreased when the message content felt lengthy, predictable in its phrasing, or overly formal. For example, P8 mentioned that when the messages' length increased, they would "just skim through it," prioritizing brevity and directness. Furthermore, P1 reported a practical limitation, noting that on their older phone, longer texts were sometimes cut off entirely, rendering the latter part of the message useless.

Participants were divided over the message structure. Some found the consistent structure beneficial, adding that it made the messages "very easy to understand," and allowed them to quickly identify key information. Thus, the predictability offered a sense of familiarity and reduced their cognitive load. However, other participants found that the same consistency led to feelings of repetition and predictability. For instance, P3 highlighted that this lack of variation made it easy to "tune out" once they had become accustomed to the general structure of the messages.

Participants also reported that the messages’ tone of voice adversely impacted their engagement. The messages felt impersonal, and P3 explained, \textit{"I feel like I have a tendency to ignore messages that just seem generic."} The impersonality also led participants to feel that the messages were AI-generated, which promoted bias against the system. For example, P7 stated that a message from an LLM "wouldn’t give [them] the same strength of confidence as … a real person." P8 also expressed skepticism about AI-based systems and losing "that sense of real-life community" from overly-relying on these tools.

\section{Discussion}
Our findings answer our research question by highlighting that university students envision LLMs to pick out stressful events from calendar data, and adapt them to their context using a concise and colloquial tone. We discuss how these insights relate to existing literature and draw on them to generate design recommendations for LLM-based tools leveraging calendar data to manage stress.

\subsection{Need for Structured Guidance to Address Personalization}
Participants reported that receiving messages for low-stakes events, or the system misinterpreting their event, led to impersonal messages that decreased their engagement. While LLMs are typically celebrated for their ability to enable dynamic personalization \cite{kumar2024large, bhattacharjee2024explains}, our findings highlight an important limitation: LLMs require structure to interpret events. To elaborate, they need well-organized contextual information to be able to differentiate low-stakes events from stressful events, and adapt themselves to each event. Future studies can complement calendar data with a sequence of targeted questions that can be asked to participants during idle time \cite{poole2013hci}. These multi-series questions can focus on specific aspects of participants’ thoughts about an upcoming event, allowing for a more structured presentation of thoughts so LLMs can capture nuances they might have originally missed. However, it is important to consider the timing and frequency of these questions in order to maintain a productive balance between personalization and participant burden. 

\subsection{Fostering Trust and Disclosure through Colloquial Tone}
Participants found that the messages felt overly formal and LLM-driven. This often introduced implicit biases against the system, negatively affecting participants’ willingness to engage or disclose personal information due to diminished trust. This aligns with existing literature, which states that awareness of AI involvement in mental health interventions can trigger skepticism or discomfort among participants, potentially undermining intervention efficacy \cite{bhattacharjee2024understandingMSR, bhattacharjee2024explains}. Interestingly, Kumar et al. \cite{kumar2024large} demonstrated that when LLM agents adopt more socially relatable communication patterns, such as engaging in casual small talk, they may become more productive at fostering user engagement. As a result, future work should instruct LLMs to mirror the participant’s conversational tone, which can enhance trust and participant disclosure, thereby strengthening the effectiveness of LLM-driven stress interventions.

\section{Limitations and Future Directions}
We note that our participant pool was relatively small ($N=8$) and recruited via snowball sampling. This may have introduced sampling bias. Engaging with more participants would provide a broader perspective on how users respond to context-aware interventions tied to their calendar events, potentially revealing novel challenges and preferences. 

Additionally, our study focused on short-term engagement with the tool. As a result, our findings are best framed as an initial probe into how users engage with and perceive a calendar-integrated, LLM-based stress intervention. It is important to note, however, that users' stated preferences might diverge from actual user experience in prolonged deployment. With suitable researcher oversight and ethical considerations, future studies could include longitudinal deployments that integrate our design implications to empirically assess their helpfulness and user acceptance.

\section{Conclusion}
Existing stress interventions frequently overlook the contextual complexity and personal nuances of students’ schedules. LLMs present a promising opportunity to address this gap by adapting support to natural language inputs and calendar data. Yet, the expectations and design considerations surrounding such integrations remain underexplored. To help fill this gap, we conducted a one-week deployment of a calendar-integrated, LLM-enhanced messaging probe with eight university students, followed by in-depth interviews.

Our findings highlight the importance of prioritizing messages around genuinely stressful events and using a concise yet colloquial tone. These insights led us to propose key design considerations for future systems, including using structured questions to supplement vague calendar entries and adopting a tone that fosters user trust and emotional disclosure. This study significantly contributes to understanding initial user expectations and reactions to calendar-aware, LLM-enhanced stress support. We believe these recommendations can inform the development of LLM-based stress management tools that are personally meaningful, and sensitive to users’ lived experiences.

\begin{acks}

This work was supported by grants from the National Institute of Mental Health (K01MH125172, R34MH124960), the Office of Naval Research (N00014-18-1-2755, N00014-21-1-2576), the Natural Sciences and Engineering Research Council of Canada (RGPIN-2019-06968), and the National Science Foundation (2209819). In addition, we acknowledge a gift from the Microsoft AI for Accessibility program to the Center for Behavioral Intervention Technologies that, in part, supported this work (\url{http://aka.ms/ai4a}). Ananya Bhattacharjee would also like to acknowledge support from the Inlight Research Fellowship, the Schwartz Reisman Institute Graduate Fellowship, and the Wolfond Scholarship in Wireless Information Technology.

\end{acks}

\bibliographystyle{ACM-Reference-Format}
\bibliography{main}

\end{document}